\documentclass[letterpaper, 10 pt, conference]{ieeeconf} 
\IEEEoverridecommandlockouts                              
\usepackage{mathtools}
\usepackage{amssymb}
\usepackage{bbold}
\usepackage[yyyymmdd,hhmmss]{datetime}
\usepackage{bm,upgreek}
\usepackage{amssymb}  
\usepackage{color}
\mathtoolsset{showonlyrefs,showmanualtags}
\usepackage{dsfont}
\usepackage{booktabs}
\usepackage[final]{graphicx}
\usepackage{nth}

\DeclareMathOperator{\Binopdf}{Bino}
\newcommand{\ID}[1]{ \mathbb{1} \left(#1 \right) }  
\newcommand{\Exp}[1]{\mathbb{E}\big[ #1\big]}
\newcommand{\Prob}[1]{\mathbb{P} \big(#1\big)}       
 
\newcommand{\Mspace}{\mathcal{M}_n} 
\usepackage[standard]{ntheorem}
\newtheorem{Assumption}{Assumption}
\newtheorem{Problem}{Problem}

\title{\LARGE \bf
A Mean-Field Team Approach  to  Minimize  the Spread of Infection in a Network}

\author{Jalal Arabneydi and Amir G. Aghdam
\thanks{This work has been supported in part by the Natural Sciences and Engineering Research Council of Canada (NSERC) under Grant RGPIN-262127-17, and in part by Concordia University under Horizon Postdoctoral Fellowship.}  
\thanks{Jalal Arabneydi and Amir G. Aghdam are with the  Department of Electrical and Computer Engineering, 
        Concordia University, 1455 de Maisonneuve Blvd. West, Montreal, QC, Canada, Postal Code: H3G 1M8.  Email: {\tt\small jalal.arabneydi@mail.mcgill.ca},        
        {\tt\small aghdam@ece.concordia.ca}}%
}

\begin{document}
\maketitle

\vspace*{-5cm}{\footnotesize{Proceedings of American Control Conference, 2019.}}
\vspace*{4.3cm}

\thispagestyle{empty}
\pagestyle{empty}
\begin{abstract}
In this paper,  a stochastic dynamic control strategy is presented to  prevent the spread of an infection over a homogeneous network. The infectious process is persistent, i.e.,  it continues to  contaminate the network once it is established. It is assumed that there is a finite set of network management options available such as  degrees of nodes and   promotional plans  to minimize the number of infected nodes while taking  the  implementation cost into account. The  network is modeled by an exchangeable controlled Markov chain, whose  transition probability matrices  depend on three parameters:  the selected network management option,  the state of the infectious process, and the empirical distribution of infected nodes (with not necessarily a linear dependence). Borrowing  some techniques from mean-field team theory the  optimal strategy  is obtained for any finite number of nodes using  dynamic programming decomposition and the convolution of some binomial probability  mass functions.   For  infinite-population networks,   the optimal solution is described by  a Bellman equation. It is shown that  the infinite-population strategy is  a meaningful  sub-optimal solution  for finite-population networks  if a  certain condition holds. The theoretical results are verified by an  example  of  rumor control     in   social networks.
\end{abstract}

\section{Introduction}

Networks are ubiquitous in today's world,  connecting people and organizations in various ways to improve the quality of day-to-day life in terms of, for example,  health services~\cite{anderson2013population},   consumers demand~\cite{shy2011short}, energy management~\cite{Pagani2013},  and social activities~\cite{easley2010networks},  to name only a few. There has  been a growing interest in the literature recently on network analysis, and in particular, on enhancing network reliability and security~\cite{sha2016security,friedberg2015combating}. This problem has been on the spotlight ever since the dramatic influence of social media on public opinion was observed in a number of major events.


Controlling the spread of undesirable phenomena such as disease and misinformation over a network is an important problem for which different approaches are proposed in the literature~\cite{pastor2015epidemic,nowzari2016analysis}. The dynamics of an infection propagating in a network of $n$ nodes, where each node has a binary state (susceptible and infected),  can be modeled by a Markov chain with a $2^n \times 2^n$ transition probability matrix. Since the computational complexity of such a model is exponential with respect to $n$, mean-field theory has proved to be effective in approximating a large-scale dynamic network by an infinite population one. For this purpose, the dynamics of the probability distribution of infected nodes can be described by a differential equation (called diffusion equation)~\cite{kephart1992directed,nekovee2007theory}.

In the analysis of infection spread,  the main objective is to study the dynamics of  the states of  nodes, specially after a sufficiently long time, in order to determine  the rate of convergence  to the steady state~\cite{van2009virus,wang2003epidemic,HassibiCDC2015}. It is shown in~\cite{van2009virus}   that if the rate of spread of infection over the rate of cure  is less than the inverse of the largest eigenvalue of the adjacency matrix, the infinite-population network reaches to an absorbing state in which all states are healthy, i.e., the infection is eventually cleared.

In the control of infection spread, on the other hand, the objective is to derive the transition probabilities such that a prescribed performance index, which is a function of implementation cost and the number of infected nodes, is minimized~\cite{morton1974optimal,khanafer2014optimal,eshghi2016optimal,di2016optimal,nowzari2017optimal}. This problem is computationally difficult to solve, in general. However, in the special case when the diffusion equation and cost function have certain structures, the optimal strategy can be obtained analytically. For example, in~\cite{morton1974optimal,khanafer2014optimal,eshghi2016optimal} it is assumed that the network dynamics  and cost are linear in control action (that is immunization or curing rates),  which  leads to bang-bang  control strategy.   The interested reader is referred to~\cite{di2016optimal,nowzari2017optimal} for more details on optimal resource allocation  methods.

This paper studies  the  optimal control  of a network consisting of an arbitrary number of  nodes that are influenced (coupled)  by the empirical distribution of infected nodes (such couplings are not  necessarily linear). The infectious process is assumed to be persistent, in the sense that the infection does not disappear after the initial time. In contrast to the papers cited in the previous paragraph which consider a continuous action set, in this paper it is assumed that there is a limited number of resources available, which means that the action set is finite. In addition, we raise a practical   question that when the solution of an   infinite-population network constructs a meaningful approximation for the finite-population one. Inspired by  existing  techniques for mean-field teams~\cite{arabneydi2016new,JalalCDC2017,JalalACC2018,Jalal2018fault,JalalCCECE2018,JalalCDC2018}, we  first compute the optimal solution of a finite-population network for the case  where  the empirical distribution of infected nodes is  observable. Next,   we  derive an infinite-population Bellman equation that requires no observation of  infected nodes, and  identify a stability condition  under which the solution of the infinite-population network constitutes  a near-optimal solution  for the finite-population one. 

The paper is structured as follows. In Section~\ref{sec:prob} the problem is formulated and the objectives are subsequently described. The optimal control strategies, as the main results of the paper, are derived on micro and macro scales in Section~\ref{sec:theoretical}. An illustrative example of a social network is presented in Section~\ref{sec:example}. The results are finally summarized in Section~\ref{sec:conclusions}.


\section{Problem Formulation}\label{sec:prob}
\subsection{Notational convention}
Throughout this article,  $\mathbb{R}$   is  the   set of real numbers  and $\mathbb{N}$   is   the set of natural numbers.  For any $n \in \mathbb{N}$,  let $\mathbb{N}_n$ and  $\Mspace
$  represent  finite sets $\{1,\ldots,n\}$ and  $\{0,\frac{1}{n}, \frac{2}{n},\ldots,1\}$, respectively,  and  $x_{1:n}$   denote   the vector $(x_1,\ldots,x_n)$.  In addition,  $\mathbb{E}[\boldsymbol \cdot]$, $\mathbb{P}(\boldsymbol \cdot)$ and $\ID{\boldsymbol \cdot }$   refer  to  the expectation,  probability and indicator operators, respectively.  For any $n \in \mathbb{N}$ and $p \in [0,1]$, $\Binopdf{(\boldsymbol \cdot, n,p)}$  is the binomial probability distribution function  of $n$ binary trials with success probability~$p$.

\subsection{Model}
Consider a population of $n \in \mathbb{N}$ homogeneous  users   that are exposed to an   infectious process (e.g., disease or fake news).  Let  $x^i_t \in \{S,I\}$  be the state of  user $i \in \mathbb{N}_n$ at time $t \in \mathbb{N}$, where $S$ and $I$  stand  for ``susceptible'' and ``infected'', respectively.  Denote  by  $m_t \in \Mspace$   the empirical distribution of the infected users at time $t \in \mathbb{N}$, i.e., $m_t=\frac{1}{n} \sum_{i=1}^n \ID{x^i_t=I}$.

\subsubsection{\textbf{Resources}} Let $\mathcal{U}$  denote the set of finite options  available to  the network manager  (e.g., a company or a government).  The objective  of the network manager   is  to  minimize the effect of the  infectious process on  the  users by  employing the available  options effectively.    For instance,  one possible option is the degree of   nodes and by varying the degree (i.e.,  topology),  the spread of an infection can be impeded. Alternatively, the option  may be an action plan such as vaccination or health promotion,   influencing the rates of  infection and cure. Denote by  $u_t \in \mathcal{U}$  the option  taken by the network manager  at time $t \in \mathbb{N}$.

\subsubsection{\textbf{Infectious process}} Let $z_t \in \mathcal{Z}$  be  the state of an infectious  process  at time $t \in \mathbb{N}$, where $\mathcal{Z}$ is a finite set consisting of all  possible states.  Denote by $\Prob{z_{t+1} \mid z_t,u_t}$ the  transition probability according to which  state $z_t \in \mathcal{Z}$ transits  to state $z_{t+1} \in \mathcal{Z}$ under   option $u_t \in \mathcal{U}$, $\forall t \in \mathbb{N}$.   Note that  the level of persistence of the  infectious process  is  incorporated in  the above transition probability matrix.

\subsubsection{\textbf{Dynamics of users}}  Suppose  that the state of user $i$ is susceptible, the state of the infectious process is $z_t$,  option $u_t$ is  chosen and  the number of infected users is $nm_t$, $t \in \mathbb{N}$.  Then,    user $i$ becomes infected   with the following probability:
\begin{equation}\label{eq:f0}
\Prob{x^i_{t+1}=I \mid x^i_t=S,  u_t,m_t,z_t}:= f^0(u_t,m_t,z_t),
\end{equation}
where $f^0:   \mathcal{U} \times \Mspace  \times  \mathcal{Z} \rightarrow [0,1]$. In addition,   when the state of user $i$ is infected, it  changes to  susceptible according to the following  probability:
\begin{equation}\label{eq:f1}
\Prob{x^i_{t+1}=S \mid x^i_t=I,  u_t,m_t,z_t}:= f^1(u_t,m_t,z_t),
\end{equation}
where $f^1:  \mathcal{U} \times \Mspace \times  \mathcal{Z}  \rightarrow [0,1]$.   It is to be noted that the  network topology  is  implicitly described in transition probabilities~\eqref{eq:f0} and~\eqref{eq:f1}.

\subsubsection{\textbf{Per-step cost}}  Let  $c(u,m,z) \in \mathbb{R}_{\geq 0}$  be the cost associated with   implementing option $u \in \mathcal{U}$ when the empirical distribution  of the infected users is $m \in \Mspace$ and the state of the infectious process is $z \in \mathcal{Z}$.  For practical purposes, the per-step cost function is considered to be  an increasing function  of  the empirical distribution  of the infected users, i.e.,  the more  infection, the higher  cost.

 At any  time $t \in \mathbb{N}$,   the network manager   chooses its option  according to the control law   $g_t: (\mathcal{M}_n \times \mathcal{Z})^t \rightarrow \mathcal{U}$ as follows:
\begin{equation}
u_t=g_t( m_{1:t},z_{1:t}), \quad t \in \mathbb{N}.
\end{equation}
Note that $g:=\{g_1,g_2,\ldots\}$ is  the  strategy  of the network manager. 

\subsection{Problem statement}
\begin{Assumption}\label{assump:primitive}
The transition  probabilities and  cost function are time-homogeneous. In addition,   the underlying  primitive random variables of users  as well as  the infectious process are   mutually independent  in both space and time.  Furthermore,    the primitive random variables of users are  identically  distributed.
\end{Assumption}

Given a discount factor $\beta \in (0,1)$, define the total expected discounted cost:
 \begin{equation}
 J_n(g)=\mathbb{E}^g \left[\sum_{t=1}^\infty \beta^{t-1}  c(u_t,m_t,z_t) \right],
 \end{equation}
 where the above cost function depends on the choice of strategy $g$ and the number of users $n$.
\begin{Problem}\label{prob1}
Find an optimal  strategy $g^\ast$ such that for  any strategy $g$,
\begin{equation}
J^\ast_n:=J_n(g^\ast) \leq J_n(g).
\end{equation}
\end{Problem}

\begin{Problem}\label{prob2}
Find a sub-optimal  strategy $\tilde{g}:=\{\tilde g_1,\tilde g_2,\ldots\}$,  $\tilde  g_t: \mathcal{Z}^t \rightarrow \mathcal{U}$, $t \in \mathbb{N}$, such that  its performance converges to the optimal performance of the infinite-population as the number of users increases, i.e., 
\begin{equation}
|J_n(\hat g) - J_\infty^\ast| \leq  \varepsilon(n),
\end{equation}
where $ \lim_{n \rightarrow \infty} \varepsilon(n)=0$.
\end{Problem}

\section{Theoretical results}\label{sec:theoretical}
Prior to solving  Problems 1 and 2, it is  necessary to  understand  the dynamics of  the empirical distribution of  the  infected users,  and more importantly,  the way  it evolves  over  time according to   each  option  of the network manager and  state of the infectious process.  To this end,   the following theorem  is needed.
\begin{Theorem}\label{thm1}
Let Assumption~\ref{assump:primitive} hold. Given any $m_t \in \mathcal{M}$, $u_t \in \mathcal{U}$ and $z_t \in \mathcal{Z}$, $t \in \mathbb{N}$, the transition probability matrix of the empirical distribution of  the infected users is characterized as:
\begin{align}
&\Prob{m_{t+1} | m_t=0,  u_t, z_t} \hspace{-.1cm}= \hspace{-.1cm}\Binopdf(nm_{t+1}, f^0(u_t, m_t,z_t), n),\\
 &\Prob{m_{t+1} |m_t=1,u_t, z_t} \hspace{-.1cm}=\hspace{-.1cm} \Binopdf(nm_{t+1}, 1-f^1(u_t, m_t,z_t), n),\\
& \Prob{m_{t+1} |m_t  \hspace{-.1cm} \notin \hspace{-.1cm}  \{0,\hspace{-.05cm}1\},u_t, z_t} \hspace{-.1cm}= \hspace{-.1cm}\Big(\hspace{-.1cm} \Binopdf(\boldsymbol \cdot, f^0(u_t, m_t,z_t), n\hspace{-.05cm}-\hspace{-.05cm}nm_t) \\
 & \quad * \Binopdf(\boldsymbol \cdot, f^1(u_t, m_t,z_t), nm_t)\Big)(nm_{t+1}).
\end{align}
\end{Theorem}
\begin{proof}
The proof proceeds in three steps. In the first step, suppose $m_t=0$, i.e. $x^i_t=S, \forall i \in \mathbb{N}_n$. In such a case, $nm_{t+1}=\sum_{i =1}^n \ID{x^i_{t+1}=I}$ is a random variable consisting of $n$ i.i.d.  Bernoulli random variables with the success probability $f^0(u_t,m_t,z_t)$. In the second step, suppose   $m_t=1$, i.e. $x^i_t=I, \forall i \in \mathbb{N}_n$. Therefore, $nm_{t+1}=\sum_{i =1}^n \ID{x^i_{t+1}=I}$ is a random variable consisting of $n$ i.i.d.  Bernoulli random variables with the success probability $1-f^1(u_t,m_t,z_t)$.  In the  last step, suppose  that $m_{t} \notin \{0,1\}$.  Then,  $nm_{t+1}=\sum_{i =1}^n \ID{x^i_{t+1}=I}$  is  the sum of two  independent random variables, where the first one  is comprised of $n-nm_t$ i.i.d. Bernoulli random variables with the success probability $f^0(u_t,m_t,z_t)$ while the second one  is comprised of  $nm_t$ i.i.d. Bernoulli random variables with the success probability $1-f^0(u_t,m_t,z_t)$. The proof is now  complete, on noting that the probability  mass function of two independent random variables is the convolution of their probability mass functions. $ \hfill \blacksquare$
\end{proof}

\begin{Theorem}\label{thm2}
Let Assumption~\ref{assump:primitive} hold. Problem~\ref{prob1} admits an optimal   stationary strategy  characterized    by the following  Bellman equation: for any $m \in \Mspace$ and $z \in \mathcal{Z}$,
\begin{multline}\label{eq:bellman-1}
V(m,z)=\min_{u \in \mathcal{U}} \Big( c(u,m,z) +\beta \big( \sum_{z^+ \in \mathcal{Z}}  \sum_{ m^+  \in \mathcal{M}}    \\
 \Prob{ z^+ \mid z,u}  \Prob{ m^+ \mid m, u,z} V(m^+,z^+)\big) \Big).
\end{multline}
Let $g^*$ be a minimizer of the above  Bellman equation; then  the optimal action  at time $t \in \mathbb{N}$ is given by  $u^\ast_t=g^\ast(m_t,z_t)$.
\end{Theorem}
\begin{proof}
From the proof of Theorem~\ref{thm1} and the fact that the infectious process evolves in a Markovian manner with a  transition probability  independent of  the states of  users, it follows that:
\begin{multline}\label{eq:proof-markov}
\Prob{m_{t+1},z_{t+1} \mid m_{1:t},z_{1:t},u_{1:t}}= \Prob{z_{t+1}\mid z_t,u_t} \\
\times  \Prob{m_{t+1}|m_t,u_t,z_t},
\end{multline} 
where the left-hand side of eqaution~\eqref{eq:proof-markov} does not depend on the  control laws $g_{1:t}$. Hence, one can find the optimal solution of Problem~\ref{prob1}  via  the  dynamic programming principle~\cite{Bertsekas2012book}, and  this leads  to   the Bellman equation~\eqref{eq:bellman-1}. $\hfill \blacksquare$
\end{proof}
According to Theorem~\ref{thm2}, the optimal strategy does not depend on the history of  infected users and infectious process, i.e.,  it is sufficient  to know     the current values in order to optimally control the network.
\begin{remark}
The cardinality of the  space of the Bellman equation~\eqref{eq:bellman-1}, i.e. $\Mspace$, is linear in the number of users $n$.
\end{remark}

In the special case of $n=\infty$,  the probability mass function of $m_{t+1}$ becomes a Dirac  measure. In such a case, there is no loss of optimality  in restricting attention to the dynamics  of the controlled differential equations. More precisely, define 
\begin{equation}
 p_t:= \lim_{n \rightarrow \infty } \frac{1}{n} \sum_{n=1}^\infty \ID{x^i_t=I}.
\end{equation} 
 According to~\cite[Lemma 4]{JalalCDC2017}, the following equality holds with probability one for  any  trajectory $z_{1:\infty}$,
\begin{equation}\label{eq:macro-dynamics}
p_{t+1}= (1-p_t)f^0(u_t,p_t,z_t) + p_t (1-f^1(u_t,p_t,z_t)),
\end{equation}
with $p_1=\Exp{m_1}$.  By  incorporating  the macro-scale (infinite-population)  dynamics~\eqref{eq:macro-dynamics} into the Bellman equation~\eqref{eq:bellman-1}, one  arrives at  the following Bellman equation: 
\begin{multline}\label{eq:bellman-2}
\tilde V(p,z)=\min_{u \in \mathcal{U}} \Big( c(u,p,z) + \beta  \sum_{\tilde z \in \mathcal{Z}}       \Prob{\tilde z \mid z,u} \\
\times  \tilde V\big((1-p)f^0(u,p,z) + p (1-f^1(u,p,z)),\tilde z\big) \Big),
\end{multline}
for any $p \in [0,1]$ and $z \in \mathcal{Z}$. Let  $\tilde g: [0,1] \times \mathcal{Z} \rightarrow \mathcal{U}$ be a minimizer of the right-hand side of equation~\eqref{eq:bellman-2}, and  define the following action at time $t$:
\begin{equation}\label{eq:tilde_g}
\tilde u_t:=\tilde g(p_t,z_t).
\end{equation}
 Notice  that $p_{1:t+1}$ is  a stochastic process adapted to the filtration $z_{1:t}$ for any $t \in \mathbb{N}$, i.e., 
 \begin{equation}\label{eq:macro-dynamics-1}
p_{t+1}= (1-p_t)f^0(\tilde g(p_t,z_t),p_t,z_t) + p_t (1-f^1(\tilde g(p_t,z_t),p_t,z_t)).
\end{equation}
 To establish the convergence result, the following  assumptions are imposed on the model. 

 \begin{Assumption}\label{assumption:cost}
 There exist  positive constants $k^0,k^1,k^c \in \mathbb{R}_{\geq 0}$ such that given any $u \in \mathcal{U}$, $z \in \mathcal{Z}$ and $m^1,m^2  \in [0,1]$, 
\begin{align}
|f^0(u,m^1,z)-f^0(u,m^2,z)| \leq k^0 |m^1-m^2|,\\
|f^1(u,m^1,z)-f^1(u,m^2,z)| \leq k^1 |m^1-m^2|, \\
|c(u,m^1,z)-c(u,m^2,z)| \leq k^c |m^1-m^2|.
\end{align}
 \end{Assumption} 
\begin{Assumption}\label{assumption:inf}
The parameters introduced in Assumption~\ref{assumption:cost} satisfy the inequality $\max(k^0,k^1) <\dfrac{1}{\beta}$.
\end{Assumption}
 \begin{Theorem}\label{thm3}
 Let Assumptions~\ref{assump:primitive},~\ref{assumption:cost} and~\ref{assumption:inf} hold. Then,  $\tilde g$ is a sub-optimal strategy for Problem~\ref{prob2}, i.e.:
 \begin{equation}
|J_n(\tilde  g) - J_\infty^\ast |\leq   \frac{k^c}{(1-\beta)(1-\beta \max(k^0,k^1))}  \mathcal{O}(\frac{1}{\sqrt n}),
\end{equation}
where $\lim_{n \rightarrow \infty}  \mathcal{O}(\frac{1}{\sqrt n})=0$.
 \end{Theorem}
 \begin{proof}
Let $\tilde m_t$ denote the empirical distribution of the infected users under strategy $\tilde g$ at time $t \in \mathbb{N}$.  For ease of display,  let function $\phi: \mathcal{U} \times [0,1] \times \mathcal{Z} $  denote the dynamics~\eqref{eq:macro-dynamics}, i.e.   $p_{t+1}=\phi(u_t,p_t,z_t), t \in \mathbb{N}$.  For  any  $\tilde m_t$, $p_t$ and $z_t$ at time $t \in \mathbb{N}$,   the following inequality holds as a result of   the triangle inequality, monotonicity of the expectation function, Assumptions~\ref{assump:primitive} and~\ref{assumption:cost}, and equations~\eqref{eq:macro-dynamics},~\eqref{eq:tilde_g} and~\eqref{eq:macro-dynamics-1}:
\begin{align}\label{eq:macro-proof1}
\mathbb{E}|\tilde m_{t+1} &- p_{t+1}|=\mathbb{E}|\tilde m_{t+1} - \phi(\tilde g(p_t,z_t),p_t,z_t)| \nonumber \\
 & \leq  \mathbb{E}|\tilde m_{t+1} - \phi(\tilde g(p_t,z_t),\tilde m_t,z_t) | \nonumber \\
 & \quad +\mathbb{E}| \phi(\tilde g(p_t,z_t),p_t,z_t)- \phi(\tilde g(p_t,z_t),\tilde m_t,z_t) | \nonumber \\ 
 & \leq \mathcal{O}(\frac{1}{\sqrt{n}}) + \max(k^0,k^1) |\tilde m_t -p_t|,
\end{align}
where rate $\mathcal{O}(\frac{1}{\sqrt{n}})$ is  the rate of convergence  to the  infinite-population limit~\cite[Lemma 4]{JalalCDC2017}.
On the other hand, from    the triangle inequality, monotonicity of the expectation function, Assumptions~\ref{assump:primitive} and~\ref{assumption:cost}, and equations~\eqref{eq:bellman-2},~\eqref{eq:tilde_g} and~\eqref{eq:macro-dynamics-1}:
\begin{align}\label{eq:macro-proof2}
&J_n(\tilde  g) \hspace{-.1cm}- \hspace{-.1cm} J_\infty^\ast= \Exp{\sum_{t=1}^\infty \beta^{t-1} c(\tilde g(p_t,z_t),\tilde m_t,z_t)} - \Exp{\tilde V(p_1,z_1)} \nonumber \\
&= \Exp{\sum_{t=1}^\infty \hspace{-.05cm} \beta^{t-1} \hspace{-.05cm}c(\tilde g(p_t,z_t),\tilde m_t,z_t) \hspace{-.1cm}-\hspace{-.1cm} \sum_{t=1}^\infty  \hspace{-.05cm}\beta^{t-1} c(\tilde g(p_t,z_t),p_t,z_t)} \nonumber \\
& \leq \sum_{t=1}^\infty \beta^{t-1} k^c \Exp{|\tilde m_t -p_t|}.
\end{align}
From~\cite[Lemma 2]{JalalCDC2017},  we have that  $\Exp{|\tilde m_1-p_1|} \leq \mathcal{O}(\frac{1}{\sqrt n})$. Then,  the proof  follows from Assumption~\ref{assumption:inf} and successively using~\eqref{eq:macro-proof1} in inequality~\eqref{eq:macro-proof2}. $\hfill \blacksquare$
 \end{proof}
\begin{remark}
 It is to be noted that no continuity assumption is imposed  on the  infinite-population strategy~\eqref{eq:tilde_g} in order  to  derive Theorem~\ref{thm3}. In addition, an extra stability condition (i.e., Assumption~\ref{assumption:inf}) is needed to ensure that the infinite-population strategy is stable when applied to the finite-population network. 
 \end{remark}
Since the optimization in~\eqref{eq:bellman-2} is over an infinite space $[0,1]$,  it  is computationally difficult to find the exact solution. However, it is shown in~\cite[Corollary 1]{JalalCDC2017} that  if the optimization problem  is carried out over space $\Mspace$, the resultant solution will be  a near-optimal solution for the finite-population case  under Assumptions~\ref{assump:primitive},~\ref{assumption:cost} and~\ref{assumption:inf}.
\begin{Corollary}\label{cor1}
Let Assumptions~\ref{assump:primitive},~\ref{assumption:cost} and~\ref{assumption:inf} hold. Let also $\tilde g_n: \Mspace \times \mathcal{Z}$ be a minimizer of the quantized version of the Bellman equation~\eqref{eq:bellman-2}, as proposed in~\cite[Corollary 1]{JalalCDC2017}.  The performance of $\tilde g_n$  converges to $J^\ast_\infty$  at the rate $1 / \sqrt{n}$.
\end{Corollary}

\section{Simulations: A social network example}\label{sec:example}
Nowadays, many people  get  their  daily news via social media, where  a small piece of false information may  propagate and lead to a widespread misinformation and potentially  catastrophic consequences. As a result,   it is crucial for  network managers  as well as governments to   prevent   large-scale  misinformation on social media. Inspired by this objective,  we present a simple rumor control problem, where the goal of a network manager is to minimize the number of misinformed users in the presence of a false rumor.
\begin{figure}[t]
	\centering
	\vspace{-3cm}
	\includegraphics[width=\linewidth]{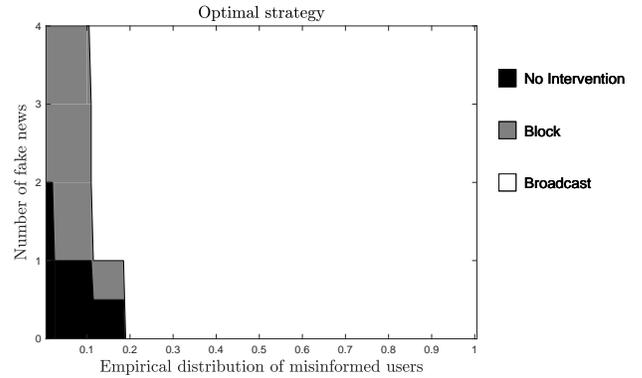} 
	\vspace{-3.5cm}
	\caption{The optimal strategy for a network of size $200$ in Example 1.}\label{fig 1}
\end{figure}

\textbf{Example 1:} Consider  $n \in \mathbb{N}$ users and  a matter of  public interest with uncertain outcome   such as an election. Let $x^i_t=S$  mean  that user $i \in \mathbb{N}_n$ at time $t \in \mathbb{N}$ is correctly informed about the topic and $x^i_t=I$ mean otherwise. Denote  by $m_t=\frac{1}{n}\sum_{t=1}^n \ID{x^i_t=I}$ the empirical distribution of the misinformed users at time $t$.

Let $z_t \in \{0,1,2,3,4\}$  be the number of  fake news that  the source of the rumor publishes   on social media at time~$t$.  The source  is assumed to be persistent, i.e.,  it  will find a way to spread the rumor unless it is constantly blocked.

 The  network manager has three options at each time instant,   i.e.  $u_t \in \{1,2,3\}$, where:
\begin{itemize}
\item  $u_t=1$ means that the network manager does not intervene;
\item  $u_t=2$ means that the network manager blocks  the source of  the  rumor, and 

\item   $u_t=3$ means that  the network manager broadcasts authenticated information  to the users and  addresses the issue publicly   for transparency. 
\end{itemize}

  Let  $w^z_t \in \{0,1\} $ be  a random one-unit   increment with  success probability $0.3$  such that at time $t \in \mathbb{N}$,
\begin{equation}
z_{t+1}=\begin{cases}
z_t+w^z_t, & z_t<4, u_t=1,3,\\
4, & z_t=4, u_t=1,3, \\
0, & u_t=2.
\end{cases}
\end{equation}

\begin{figure}[h]
	\centering
	\vspace{-.3cm}
	\includegraphics[width=\linewidth]{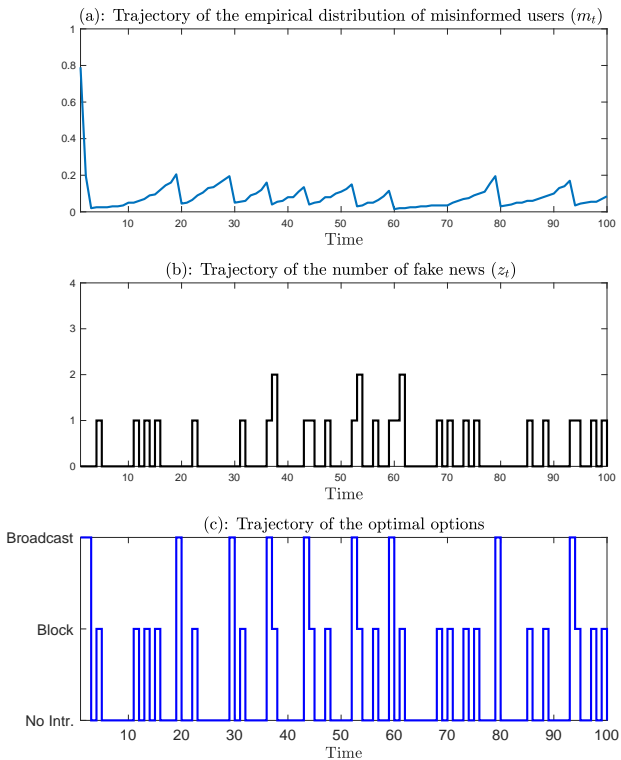} 
	\vspace{-1cm}
	\caption{Trajectory of the solution for a network of size $200$ in Example 1.}\label{fig 2}
\end{figure}

The number of fake news may be viewed as the severity level of the misinforation induced by the rumor. In this example, we have implicitly assumed that   when option ``block" is taken by the manager, the source of rumor starts producing  new fake news cautiously from zero again. The initial states  of users are identically and independently distributed   with probability mass function $(0.85, 0.15)$, where $ 0.15$ is the probability of being initially misinformed. At any time $t \in \mathbb{N}$, given the  empirical distribution of misinformed users $m_t$, the number of fake news $z_t$ and the option $u_t$ taken by the network manager,  an informed user is misled  by  the rumor  with  the following probability:
\begin{equation}
f^0(u_t,m_t,z_t)=\begin{cases}
0.2m_t(z_t+1), & u_t=1,\\
0.2m_t, & u_t=2,\\
0.1m_t^2 , & u_t=3,
\end{cases}
\end{equation}
where a larger  number of misinformed users and  fake news means  a higher probability  that a user  becomes  misinformed.  On the other hand, a misinformed  user  becomes   informed    and  convinced  by the authenticated  information  provided by  the network manager with a high probability. More precisely,
\begin{equation}
f^1(u_t)=\begin{cases}
0, & u_t=1,2,\\
0.8, & u_t=3.
\end{cases}
\end{equation}
Denote by $\ell: \mathcal{U} \times \mathcal{Z} \rightarrow \mathbb{R}_{\geq 0}$  the implementation cost  of each option, and let: 
\begin{equation}
\ell(1,z)=0, \quad \ell(2,z)=0.2z+1, \quad \ell(3,z)=5,
\end{equation}
 for any  $z \in \{0,1,2,3,4\}$. It is desired to minimize the following cost function:
\begin{equation}
\Exp{\sum_{t=1}^\infty 0.9^{t-1} \left(3.8m_t +  \ell(u_t,z_t) \right)}.
\end{equation}
To determine the optimal strategy, we  first compute the transition probability matrix in Theorem~\ref{thm1} and then solve the Bellman equation~\eqref{eq:bellman-1} in Theorem~\ref{thm2} by using the value-iteration method.  The optimal strategy for $n=200$ is displayed  in Figure~\ref{fig 1} as a function of the empirical distribution of misinformed users and the number of fake news. Under this  optimal strategy, one realization of Example 1 is depicted in Figure~\ref{fig 2}.

\begin{figure}[h]
	\centering
	\vspace{-3cm}
	\includegraphics[width=\linewidth]{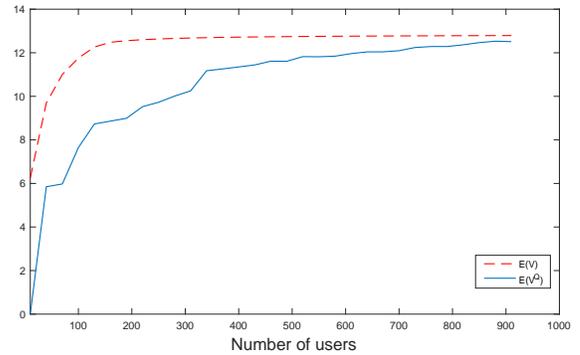} 
	\vspace{-3.2cm}
	\caption{The quantized solution converges  to the optimal solution as the number of users increases in Example 1.}\label{fig 3}
\end{figure}

To verify  the results of Theorem~\ref{thm3} and Corollary~\ref{cor1}, let the Bellman equation~\eqref{eq:bellman-2} be quantized  with a step size  $1/n$, and denote the resultant value function by $V^Q(q_1,z_1)$, where $q_1$ is the closest number in $\Mspace$ to $p_1=0.15$. Subsequently, it is shown  in Figure~\ref{fig 3} that $\Exp{V^Q(q_1,z_1)}$ converges to $J^\ast_n=\Exp{V(m_1,z_1)}$,   as $n$ increases.

\section{Conclusions}\label{sec:conclusions}
A stochastic dynamic control strategy was introduced  over a homogeneous network to minimize the spread of a persistent infection. It was shown that the exact optimal  solution can be efficiently  computed by solving a Bellman equation whose state space increases  linearly with the number of nodes. In addition, an approximate optimal solution was proposed based on the infinite-population network, where the approximation error was shown to be  upper bounded by a term that decays to zero as the number of users tends to infinity. An example of  a social network was then presented  to verify the theoretical results.

As a future research direction,  the obtained  results can be extended to partially  homogeneous networks wherein the nodes are categorized into several sub-populations of homogeneous nodes   such as low-degree and high-degree nodes.  In addition,  various  approximation methods may be used  to further alleviate the computational complexity of the proposed solutions. The development of  reinforcement learning algorithms based on  the  Bellman equations  provided in this paper can be another interesting problem for future work.

\bibliographystyle{IEEEtran}
\bibliography{Jalal_Ref}
\end{document}